\documentclass[iop]{emulateapj}

\usepackage{graphicx}
\usepackage{epsfig}
\usepackage{epstopdf}
\newcommand{\CaIIIR}{Ca~II~8542}
\newcommand{\Halpha}{H\ensuremath{\alpha}}
\newcommand{\kms}{km~s$^{-1}$}

\begin{document}
\title{On the Temporal Evolution of the Disc Counterpart of Type II Spicules in Quiet Sun}

\author{D. H. Sekse\altaffilmark{1}}
\author{L. Rouppe van der Voort\altaffilmark{1}}
\author{B. De Pontieu\altaffilmark{2}}

\affil{\altaffilmark{1}Institute of Theoretical Astrophysics,
  University of Oslo, %
  P.O. Box 1029 Blindern, N-0315 Oslo, Norway}
  
\affil{\altaffilmark{2}Lockheed Martin Solar \& Astrophysics Lab, Org.\ A021S,
  Bldg.\ 252, 3251 Hanover Street Palo Alto, CA~94304 USA}

\begin{abstract}
The newly established type II spicule has been speculated to provide enough hot plasma to play an important role in the mass loading and heating of the solar corona.  With the identification of Rapid Blueshifted Excursions (RBEs) as the on-disc counterpart of type II spicules we have analysed three different high quality timeseries with the CRisp Imaging SpectroPolarimeter (CRISP) at the Swedish Solar Telescope (SST) on La Palma and subjected to an automated detection routine to detect a large number of RBEs for statistical purposes.  Our observations are of a quiet Sun region at disc centre and we find lower Doppler velocities, 15--40~\kms, and Doppler widths, 2--15~\kms, of RBEs than in earlier coronal hole studies, 30--50~\kms\ and 7--23~\kms, respectively.  In addition, we examine the spatial dependence of Doppler velocities and widths along the RBE axis and conclude that there is no clear trend to this over the FOV or in individual RBEs in quiet Sun at disc centre.  These differences with previous coronal hole studies are attributed to the more varying magnetic field configuration in quiet Sun conditions.  Using an extremely high cadence dataset has allowed us to improve greatly on the determination of lifetimes of RBEs, which we find to range from 5 to 60~s with an average lifetime of 30~s, as well as the transverse motions in RBEs, with transverse velocities up to 55~\kms\ and averaging 12~\kms.  Furthermore, our measurements of the recurrence rates of RBEs provide important new constraints on coronal heating by spicules.  We also see many examples of a sinusoidal wave pattern in the transverse motion of RBEs with periods averaging 54~s and amplitudes from 21.5 to 129~km which agrees well with previous studies of wave motion in spicules at the limb.  We interpret the appearance of RBEs over their full length within a few seconds as the result of a combination of three kinds of motions as is earlier reported for spicules.  Finally, we look at the temporal connection between \Halpha\ and \CaIIIR\ RBEs and find that \CaIIIR\ RBEs in addition to being located closer to the footpoint also appear before the \Halpha\ RBEs.  This connection between \CaIIIR\ and \Halpha\ supports the idea that heating is occurring in spicules and contribute more weight to the prominence of spicules as a source for heating and mass loading of the corona. 
\end{abstract}

\section{Introduction}

When studying the solar chromosphere, numerous highly dynamical events can be observed taking place on very short timescales. One of these types of  events are the thin jet-like features which can be seen protruding from the solar limb, called spicules.  These jets have been observed and studied for several decades 
\citep[for extensive reviews, see][]{1968SoPh....3..367B, 
1995ApJ...450..411S, 
2000SoPh..196...79S, 
2012SSRv..tmp...65T}, 
but only in the last few years great progress in the characterisation of these events has been made.  The launch of the Hinode satellite, situated in a seeing-free environment and equipped with a large aperture, opened an unprecedented view on the solar limb with high spatial and temporal resolution.  From long-duration Ca~H timeseries obtained with Hinode, a new type of spicule was discovered 
\citep{2007PASJ...59S.655D}, 
which differentiates from the classical spicule by displaying only upward motion, shorter lifetimes and higher velocities.  These so-called type II spicules have typical lifetimes between 50-150~s, velocities between 30-110~\kms, and fade at around their maximum length 
\citep{2012ApJ...759...18P}.  
The extensive work by 
\citet{2012ApJ...759...18P} 
confirm the measurements from
\citet{2007PASJ...59S.655D} 
and contradict claims that there are no type II spicules by 
\citet{2012ApJ...750...16Z}.  

The type II spicule has been found to display a characteristic sideways swaying motion that has been regarded as a sign that the chromosphere is permeated with Alfv{\'e}nic waves with energies large enough to accelerate the solar wind and potentially heat the quiet corona 
\citep{2007Sci...318.1574D}.  
In recent years, several studies have strengthened the interpretation that type II spicules play an important role in the mass loading and heating of the corona 
\citep{2009ApJ...701L...1D, 
2011Sci...331...55D}.  
A wave analysis of high-cadence Hinode limb observations found a mix of upward propagating, downward propagating, as well as standing waves in spicules 
\citep{2011ApJ...736L..24O}.  
Whereas, 
\citet{2012ApJ...755L..11J} 
from extremely high cadence \Halpha\ wing observations, report on the sudden appearance of spicule like structures close to the limb that exhibit extreme propagation speeds.  
\citet{2012ApJ...752L..12D} 
demonstrated that the complex dynamical appearance of type II spicules must be interpreted as the result of the combination of three different types of motion: field aligned flows, swaying motions and torsional motions.

While the classical type I spicule is well understood and has been identified as the off-limb counterpart to on-disc dynamic fibrils in active regions 
\citep{2004Natur.430..536D, 
2006ApJ...647L..73H, 
2007ApJ...655..624D}, 
the formation and driver of the type II spicule are still under debate.  
In their 3-dimensional radiative magneto-hydrodynamics simulation, 
\citet{2011ApJ...736....9M} 
describe in detail the formation of jets that resemble type II spicules in that cool chromospheric plasma is ejected into the corona while being heated to temperatures that are high enough to cause apparent fading in typical chromospheric diagnostics.  The cause for these jets in the simulations is the increase in gas pressure from accelerated chromospheric plasma that is accelerated by Lorentz forces arising from flux emergence.  This triggers the ejection of plasma into the low-density corona.  Further modelling is, however, required to firmly establish the physical mechanism that drives type II spicules.

In order to improve on the simulations of spicules, high-quality observations prove to be essential when it comes to constraining the models and guide the direction of investigation.  Unfortunately, spicule properties prove to be notoriously difficult to measure at the solar limb due to the superposition of many spicules along the line of sight combined with their narrow spatial extent and significant displacement during their short lifetime. The problem of superposition can be overcome by observing their disc counterpart.  The disc counterpart of type II spicules has been identified as short-lived and rapidly moving absorption features in the strong chromospheric \Halpha\ and \CaIIIR\ spectral lines 
\citep[][Paper I]{2008ApJ...679L.167L, 
2009ApJ...705..272R}.  
\citet[][Paper II]{2012ApJ...752..108S} 
greatly expanded on the statistics of these events, commonly referred to as "Rapid Blueshifted Excursions" (RBEs).

Here we continue the work of Paper~II by addressing various properties associated with the temporal evolution of RBEs.  We use several datasets: the dual line \Halpha+\CaIIIR\ dataset from Paper~II, as well as two new \Halpha\ datasets, of which one has an extremely high cadence of 0.88~s.  This dataset allows us to properly resolve the RBEs in time so that the lifetime of RBEs can be firmly established.  The extreme temporal resolution merits further investigation of the apparent motion of some of the RBEs.  The new datasets are of quiet Sun and allow to investigate differences with coronal hole RBEs as the previous studies, Paper~I and II, are based on coronal hole target regions.

\section{Data reduction and Observations}

The observations were obtained using the CRisp Imaging SpectroPolarimeter
 \citep[CRISP,][]{2008ApJ...689L..69S} 
 instrument at the Swedish 1-m Solar Telescope 
\citep[SST,][]{2003SPIE.4853..341S} 
on La Palma.  The CRISP instrument contains a dual Fabry-P\'{e}rot interferometer (FPI) and three high-speed low-noise CCD cameras which operate at a frame rate of 35 frames per second with an exposure time of 17~ms per frame.  
Two of the cameras are located behind the FPI and a polarising beam splitter.  The third camera, which is used as an anchor channel for image processing, is positioned before the FPI\@.  All three cameras are located behind the CRISP pre-filter and are synchronised by an optical chopper.  CRISP has a field of view (FOV) of approximately 61\arcsec$\times$61\arcsec with an image scale of 0\farcs0592/pixel and allows for a fast wavelength tuning ($<$50 ms) between any two positions within the spectral region given by the spectral width of the pre-filter.  These high speed capabilities of CRISP makes it an ideal instrument for studies of the dynamic chromosphere using imaging spectroscopy.  In this study we are interested in the \Halpha\  and \CaIIIR\ lines, for which CRISP has transmission FWHM of 66~m\AA\ and 111~m\AA, and pre-filter FWHM of 4.9~\AA\ and 9.3~\AA, respectively.  
Through the use of the SST adaptive optics system 
\citep[SST,][]{2003SPIE.4853..370S} 
and the Multi-Object, Multi-Frame Blind Deconvolution image restoration technique 
\citep[MOMFBD][]{2005SoPh..228..191V}
, we obtain high spatial resolution down to the telescope diffraction limit ($\lambda/D$=0\farcs14).  Employing the cross-correlation method developed by 
\citet{2012A&A...548A.114H}, 
minimises the remaining small scale seeing signal introduced from the non-simultaneity of the narrowband CRISP images.

More information on the details of the MOMFBD restoration and further processing steps can be found in Paper~II.

On 2011 May 5 (08:31-09:02 UT), a 31-minute dataset was obtained.  The telescope was pointed at disc centre, (x,y)$\approx$(19\arcsec,5\arcsec), 
which was a quiet Sun region with some magnetic network concentrations scattered over the FOV, see Figure \ref{fig:rbe_fov}.  
This dataset was optimised for studying the fast temporal evolution of RBEs and consisted of only 4 line positions, $\pm$1032~m\AA, line core, and $-$2064~m\AA, allowing for a cadence of 0.88~s.  This extremely fast cadence allows us to resolve the temporal evolution of RBEs to much shorter time scales than previous studies.  In addition to this dataset, we acquired a time series of the same target just prior to this observation which consisted of 35 line positions symmetrically sampled up to $\pm$1290~m\AA, with 86~m\AA\ spacing, plus four additional blue wing positions going as far as $-$2064~m\AA, resulting in a cadence of 7.96~s.  This other dataset enables us to verify the validity of our detection methods, and explore differences between RBEs in quiet Sun and coronal hole from previous studies.

Finally, we analyse the dual-line \CaIIIR\ and \Halpha\ dataset from 
Paper~II.  
This is a 38-minute, 11.77~s cadence, coronal hole observation away from disc centre, (x,y)$\approx$($-$70\arcsec,508\arcsec), ($\mu$=0.84), targeting a unipolar magnetic island.  For more details and characteristics on this dataset we refer to 
Paper~II.  

\section{Method}

The RBEs were located within the datasets by the use of an automated detection algorithm, working on \Halpha\ Dopplergrams, which is  mostly identical to the one used in 
Paper~II.  
The only differences are some minor modifications allowing us to detect shorter events, $\approx$~600~km (14 pixels), than 
Paper~II 
where a threshold of 17 pixels or $\approx$~725~km was used in order to reduce the number of false positives, which resulted in a steep and possibly artificial cutoff in the lower limit of RBE lengths.  Because of the excellent quality of the data used in this study we can reduce the allowed length of detected RBEs to give us an increase in reliable detections.  We set the new lower limit of allowed RBE lengths at 14 pixels due to an exponential increase in the detection of false positives at slightly lower lengths.  We also found that using \Halpha\ Dopplergrams instead of the difference images of 
Paper~II 
gives more reliable detections, as fainter RBEs stand out more clearly, as well as a slight increase in the number of detections.

The automated detection routine determines the location of each blueshifted event in both space and time, and attempts to link these single timeframe events over time, creating multi-frame RBE chains, a process that is explained in detail in Paper~II and not changed for this study.  The very high cadence of 0.88~s of this dataset provides an excellent opportunity to improve on the previous measurements of the transverse motions and lifetimes of RBEs, which were limited by low cadence.  An average lifetime of approximately 40~s for RBEs and a cadence of the datasets on the order of 10~s, means that multi-frame RBEs were only detectable in 4-5 frames on average, and could potentially suffer from an artificial cutoff in lifetimes around 20~s, which corresponds to two frames.  In addition, the measurement of transverse motions suffered from strong constraints due to the requirement to introduce an upper limit for the transverse displacement of a multi-frame RBE from one time step to the next in order to avoid linking up single-frame RBEs into false multi-frame RBEs in crowded regions. 

Lifetime and transverse motion measurements have been improved on as compared to Paper~II, where only the duration and displacement of the chains that made up the multi-frame RBEs detected by the automated method were taken into account.  Detecting RBEs is tricky and we note that the automated detection method has a limited success rate.  To take into account that the method fails to identify all possible detections, due to too faint RBEs being ignored and too short detections being discarded, we choose to improve on the method to measure lifetimes and displacements by introducing a semi-automated method.  Here, the detection chains were used as a basis for spacetime diagrams by the use of slits, which are fixed in space, with a width of 20 pixels along the detected multi-frame RBEs.  The images of the slits are recorded far into the past and future of the RBEs and each image is averaged over the width of the slits.  On these spacetime diagrams, a manual measurement is performed, recording both the lifetime of the event as well as the transverse motion of the RBE during its lifetime.  Any multi-frame RBEs that are found to be erroneous detections based on the image in the spacetime diagram are manually removed from the sample.

Due to the intrinsic difficulty of measuring longitudinal velocities systematically as RBEs display sideways motions and seem to grow erratically, we choose to manually select a number of examples to illustrate the typical behaviour.

For the verification of multi-frame RBEs and the creation of the spacetime diagrams, we used a widget based analysis tool called CRISPEX 
\citep{2012ApJ...750...22V}, 
which allows for efficient exploration of multi-dimensional datasets.
 
\section{Results}

\subsection{Quiet Sun RBEs}
\label{sect:QS_res}

\begin{figure*}[!t]
\begin{center}
\includegraphics[width=\textwidth]{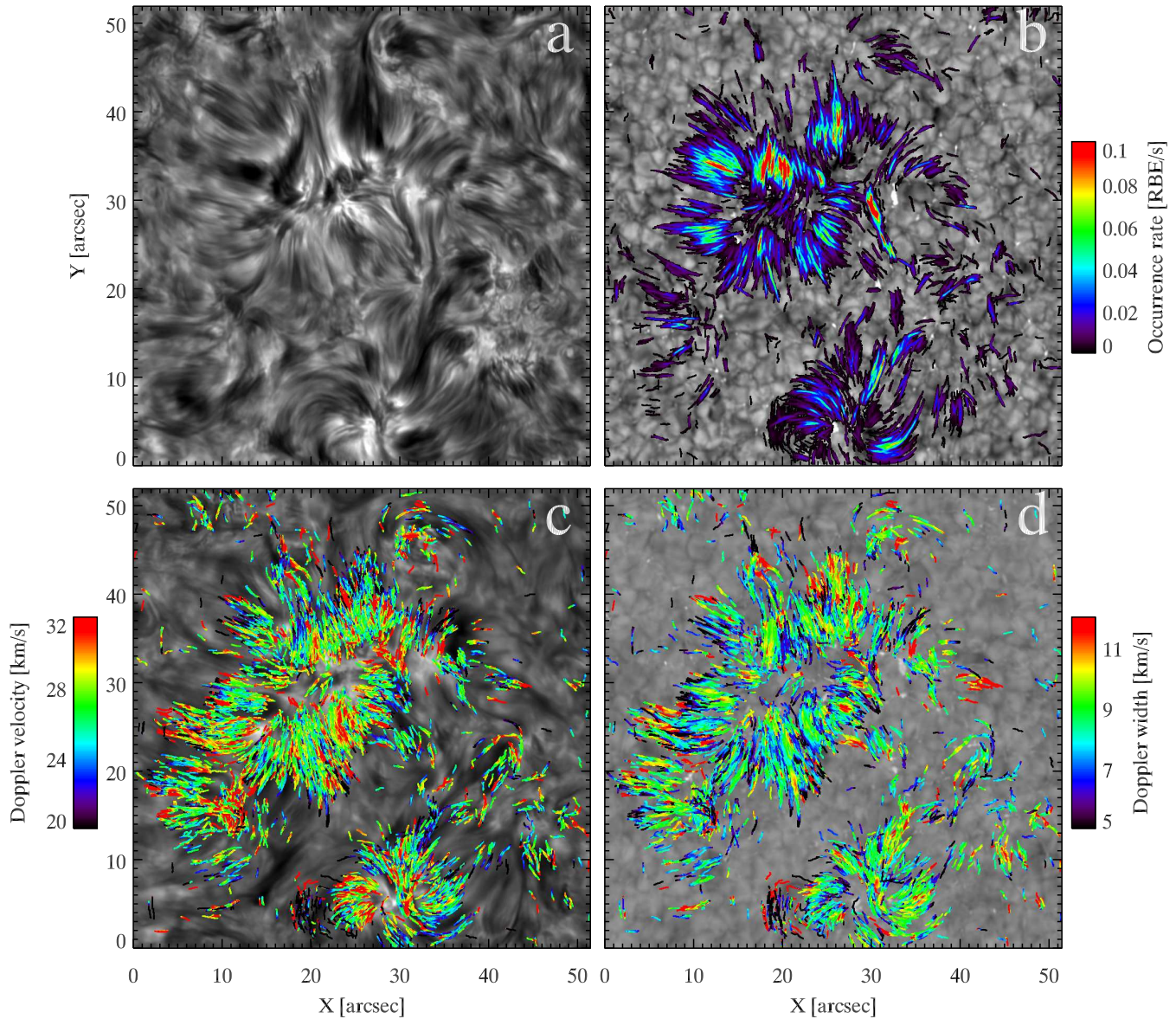}
\caption{A clean image of the FOV taken from the fast cadence scan at line centre (a).  An occurrence map of identified RBEs detected at 45~\kms\ over the whole fast cadence time series in rainbow colour scale plotted over an \Halpha\ blue wing image taken at 45~\kms\ in the fast cadence time series (b).  Doppler properties of the identified RBEs in the full scan time series as a function of their position in the FOV plotted over an image of the line centre as seen in the full scan timeseries (c) and an image of the 45~\kms\ blue wing position in the full scan dataset (d).}
\label{fig:rbe_fov}
\end{center}
\end{figure*}

\begin{figure*}[!t]
\begin{center}
\includegraphics[width=\textwidth]{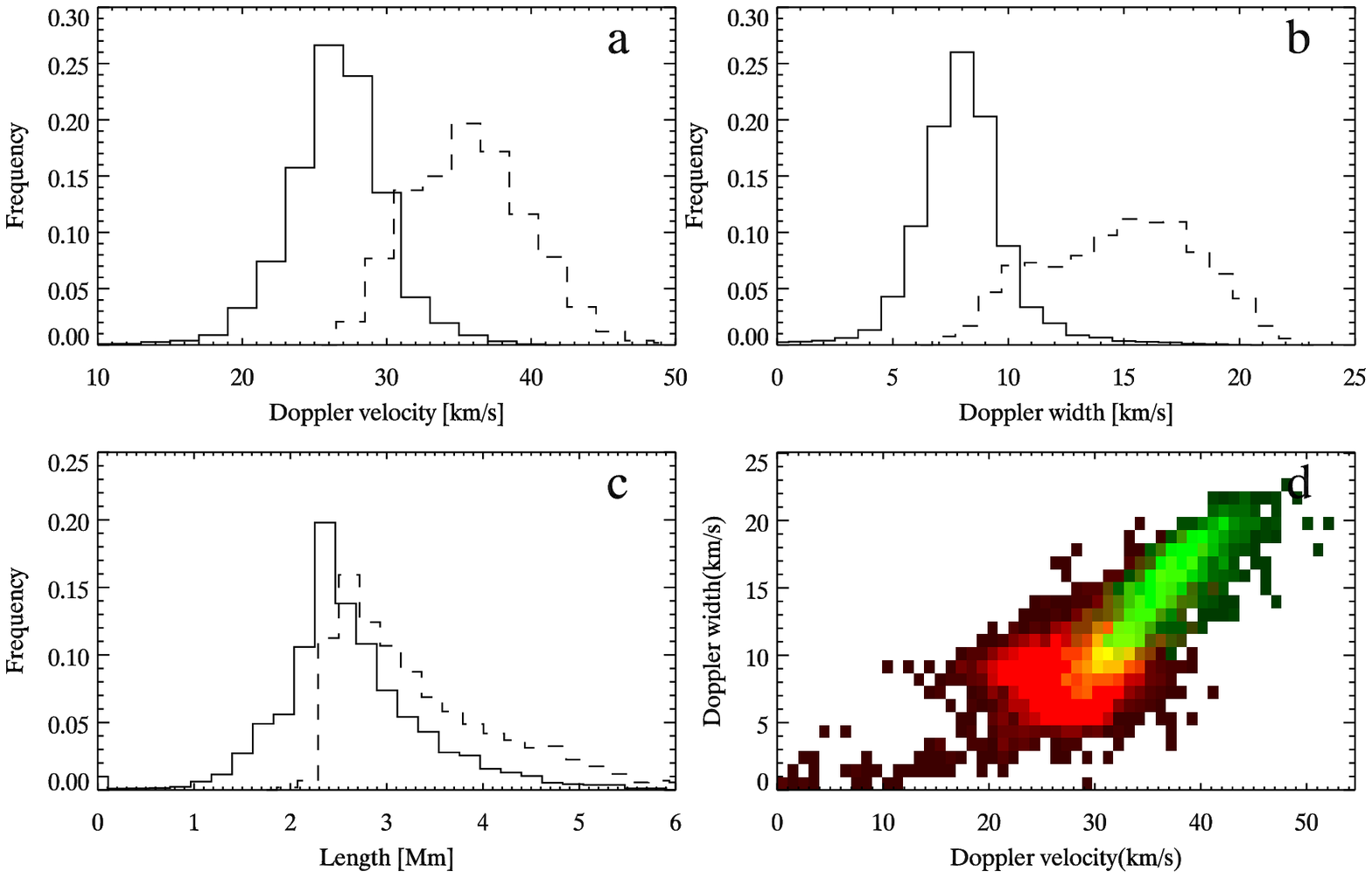}
\caption{Histograms for (a) Doppler velocity, (b) Doppler width, and (c) length of multi-frame RBEs.  Doppler velocity and Doppler width are average measures for each time step in each event.  All histograms contain data from the 5041 events that make up the multi-frame RBEs.  The dashed lines show the results of Paper~II.  
(d) is a density scatterplot of Doppler width against Doppler velocity with data from 2011 May 05 in red and Paper~II in green while yellow signifies overlap.}
\label{fig:qs_stats}
\end{center}
\end{figure*}

The two different observations of the same quiet Sun region, with different cadences and spectral samplings were searched for RBEs.  In the dataset with a cadence of 0.88~s, a total of 43077 events were detected within 31 minutes, giving a detection rate of 20.48 per frame.  The dataset with a cadence of 7.96~s produced 6795 detected events over 52 minutes with a detection rate of 17.16 per frame.

In Fig.~\ref{fig:rbe_fov} the location of every detected event for both datasets are plotted on the FOV.  Panel a) contains a clean image in \Halpha\ linecentre of the FOV seen in the fast cadence dataset while panel b has every detected event and the occurrence rate of RBEs per second from the fast cadence dataset plotted over a blue wing image from \Halpha.  The wing position used is the same as the one used when creating the detection Dopplergrams.  On average, around magnetic network where RBEs are detected and considering only those pixels that have an RBE at least once during the time sequence, the occurrence rate is 0.012 RBEs per second or one RBE every 84~s.  Averaging this number over the full 61\arcsec$\times$61\arcsec area of the FOV, the occurrence rate is 0.0033 or every pixel in the FOV sees on average one RBE every 303~s.

For the last two panels, the underlying image is the same as for panel a and b, but using the low cadence dataset.  Overplotted on these images are the positions of every detected event from the low cadence dataset along with their Doppler velocities and Doppler widths in panel c and d, respectively.

Histograms of the Doppler velocity and Doppler width of quiet Sun RBEs are displayed in Fig.~\ref{fig:qs_stats} with Doppler velocity, in panel a, showing a range from around 15~\kms\ up to 40~\kms, while Doppler widths have a range from 2--3~\kms\ up to 15~\kms.

Panel c in Fig.~\ref{fig:qs_stats} show the lengths of quiet Sun RBEs which vary between 1~Mm and 6~Mm.  For comparison the corresponding data from the coronal hole studies in 
Paper~II 
is added with dashed lines.

The scatterplot in panel d shows the distribution of the average Doppler width versus the average Doppler velocity in our quiet Sun data, in red, and is compared to the same type of scatterplot from 
Paper~II's 
coronal hole study, in green.

\subsection{Spatial Dependence of RBE Properties}
\label{sect:spatial_res}

\begin{figure}[htbp]
\begin{center}
\includegraphics[width=\columnwidth]{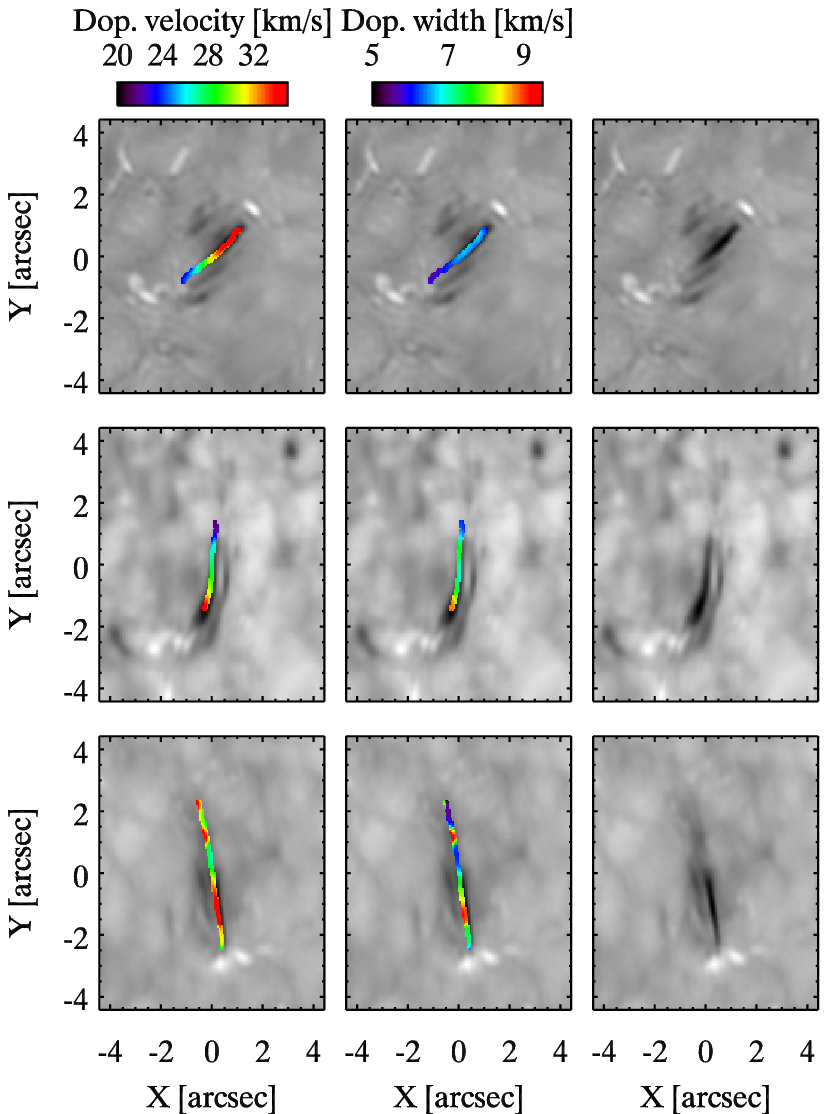}
\caption{Three examples of the variation of Doppler velocities and Doppler widths along RBEs.  Underlying images are Doppler images at the detection velocity from approximately the middle of the RBEs lifetimes.  The detected trajectories of the RBEs are plotted showing Doppler velocities in left column and Doppler widths in the middle column.  The right column contains a clean detection image of the RBEs.}
\label{fig:good_bad_ugly}
\end{center}
\end{figure}

Panels c and d of Fig.~\ref{fig:rbe_fov} show the Doppler velocity and Doppler width along the RBEs for all events located in the timeseries.  Most of the RBEs are concentrated around the network regions in the middle left and centre-bottom part of the FOV.  As these network regions can be interpreted as the footpoint regions for the RBEs, we have a general sense of the orientation for most of the RBEs.  With this general sense of orientation in mind, the maps give a rather chaotic impression with a large variation of the Doppler properties along the RBE axis.  There appears to be no systematic trend of most RBEs having increasing Doppler velocity or Doppler width towards their top end.  To illustrate this, we show three examples of the variation of Doppler properties along the RBE axis in Fig.~\ref{fig:good_bad_ugly}.  For all three examples, the footpoint (magnetic bright points) is located in the bottom of the subimage, and the RBE is extending towards the top.  The top row shows an RBE which has an increase in both Doppler velocity and width away from the footpoint.  In the middle row we see an example which displays a decrease in Doppler velocity and width away from the footpoint, while the final row contains an RBE which has a more variable distribution of Doppler velocities and widths along its length.
  
\vspace{1.5cm}
\subsection{RBE Dynamics from Temporally Resolved Data}
\subsubsection{Improved Lifetimes}
\label{sect:lifetime_res}
\begin{figure}[htbp]
\begin{center}
\includegraphics[width=\columnwidth]{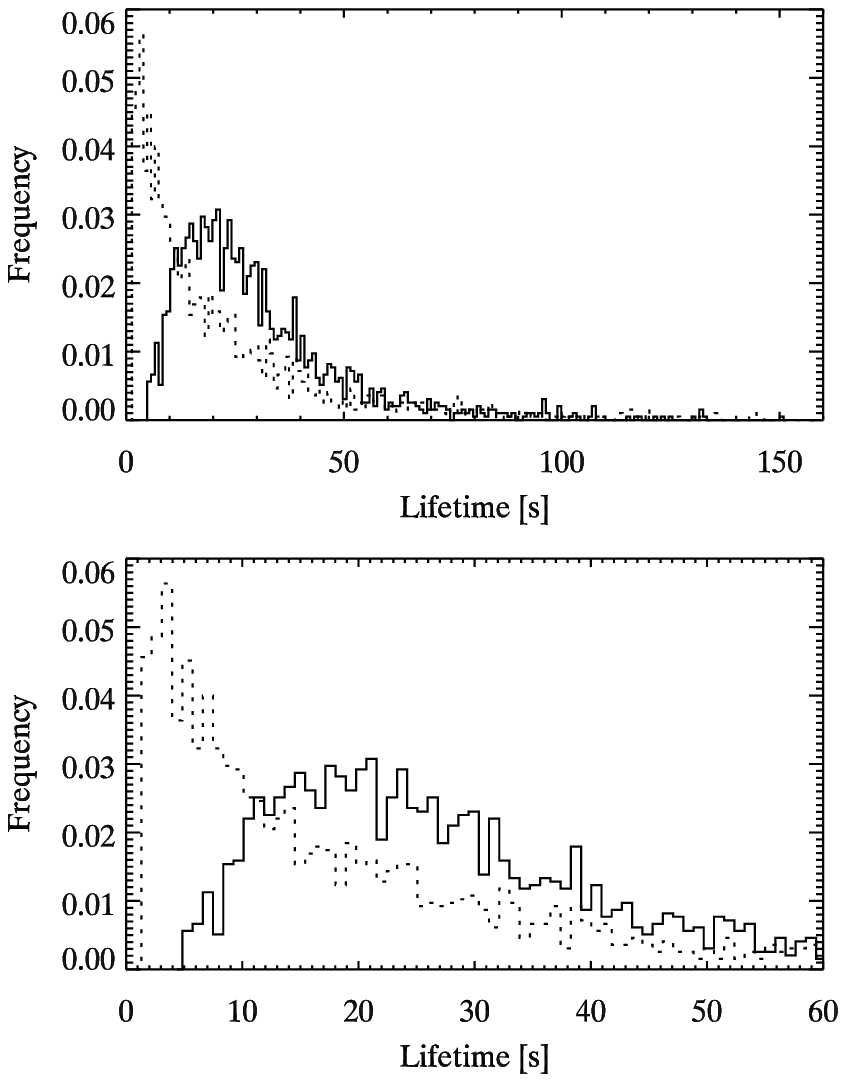}
\caption{Lifetimes of RBEs from the fast cadence dataset with the manual measurements in the solid line and the automated detection method for the same RBEs in the dashed line.  The lifetimes found by the automated detection method is added to clarify the imperfections of the detection algorithm.  There are individual events out to $\sim$150~s for the manually measured lifetimes, while the automated detection lifetimes have individual events extending as far as $\sim$350~s.  Top panel shows the entire range of lifetimes found by the manual method.  Bottom panel is a zoomed in version of the top panel.}
\label{fig:lifetime}
\end{center}
\end{figure}

By employing spacetime diagrams of multi-frame RBEs, the measurements of their lifetimes improve drastically.  Figure \ref{fig:lifetime} shows the lifetimes resulting from manual measurements on 1951 RBE spacetime diagrams in the solid line where we have a peak in the distribution around 20~s.  The shortest living RBEs have a lifetimes of 5.3~s, and are seen in only 6 frames and stand in contrast to the longest living RBE with its 150~s.

The original lifetimes found by the automated detection routine are shown in the dashed line and stand in contrast to the manually measured lifetimes.  From the manual measurements it became clear that the automated procedure failed to follow the complete lifetime for many RBEs and therefore we see a shift toward longer lifetimes of the whole distribution.  In addition, the spacetime diagrams revealed that some of the longest traced RBEs in the automated procedure, consisted in fact of several shorter lived RBEs.

 \subsubsection{Transverse Motions}
 \label{sect:transverse_res}
\begin{figure*}[!t]
\begin{center}
\includegraphics[width=\textwidth]{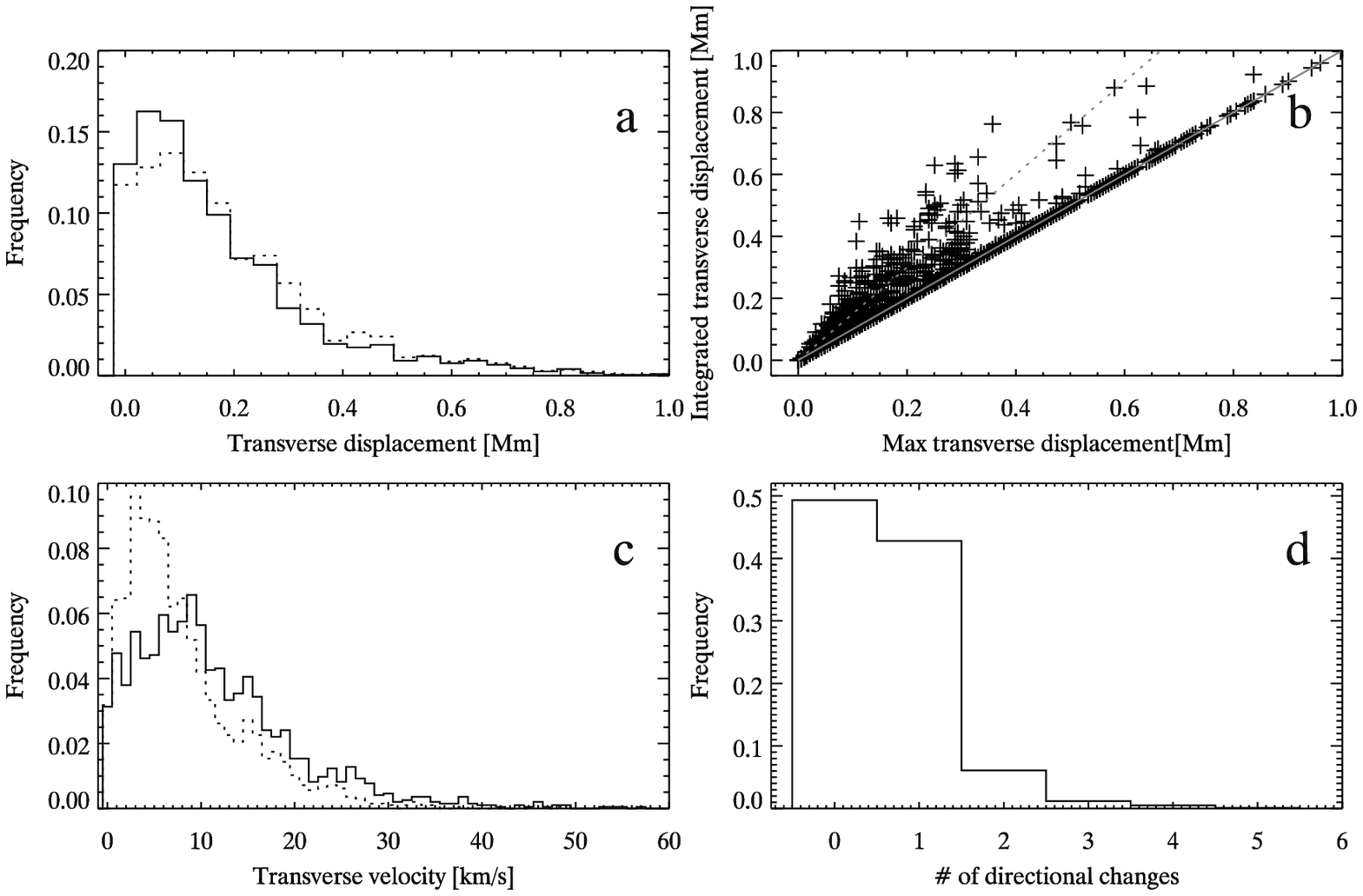}
\caption{The histogram in panel (a) shows the distance between the two extreme transverse positions during the lifetime of an RBE in the solid line plotted with the integrated or total transverse displacement covered by an RBE during its lifetime in the dashed line.  The scatter plot in panel (b) shows the relationship between the transverse distance between the two extremes of an RBE and the integrated transverse distance of the same RBE.  The solid line marks where the two measurements are equal and the dashed line marks where the integrated displacement is 50\% larger than the distance between the two extreme positions.  Panel (c) contains a histogram of the maximum transverse velocity during an RBEs lifetime in the solid line and the average transverse velocity during the lifetime of an RBE in the dashed line, while panel (d) displays the number of changes to the transverse direction an RBE is moving in.}
\label{fig:transverse}
\end{center}
\end{figure*}
 
Most RBEs display transverse displacement perpendicular to their main axis during their lifetimes.  Figure \ref{fig:transverse}a shows the distribution of the distance between the two extreme positions in the transverse motion of 1951 multi-frame RBEs, in the solid line, with an average transverse displacement of 200~km.  The distribution has a concentration towards the lower end of the scale, but there are cases of a transverse displacement above 1~Mm and the maximum transverse distance recorded is 1.23~Mm.  The dashed line in Fig.~\ref{fig:transverse}a represents the integrated transverse displacement, which will show a larger displacement for RBEs that undergo a change of direction during their lifetime as compared to the maximum displacement.  This results in an average displacement of 220~km, but the maximum recorded displacement is still 1.23~Mm.
In panel b of Fig.~\ref{fig:transverse}, the integrated transverse displacement is plotted against the maximum transverse displacement as a measure for the transverse path taken by an RBE, where the dotted line marks where the integrated transverse distance is 50\% larger than the distance between the two extreme transverse positions.  $\sim$13\% of the RBEs lie above this dotted line.  
The solid line on the other hand, marks where the two transverse displacements are equal and hence, RBEs that lie close to this line are mainly moving in one direction.  

Panel c shows the distribution of two measures of the the transverse velocity: the maximum transverse velocity of each RBE in the solid line, and the average unsigned transverse velocity over the whole lifetime of the RBE in the dashed line.  The maximum transverse velocity has a peak around 9~\kms\ and the average of the distribution is 11.77~\kms\ with highest transverse velocity recorded at 57.87~\kms. 
For the average velocities, the mean of the sample lies at 8.54~\kms\ with a peak of the distribution around 3~\kms, while the maximum average transverse velocity is found to be 49.93~\kms.

From Fig.~\ref{fig:transverse}b it is clear that RBEs that do not lie on the solid line have to display some sort of back and forth swaying motion, while those that lie above the dotted line would be more extreme cases of large back and forth swaying.
Panel d in Fig.~\ref{fig:transverse} shows the number of changes in the direction of the transverse motion displayed by the RBEs.  On average an RBE will have 0.61 directional changes, but a few RBEs are seen to have up to 5 changes to the direction of their transverse motion. 

\subsubsection{Grouping of RBEs Based on Transverse Spacetime Diagrams} 
\label{sect:grouping_res}
 
 \begin{figure*}[!t]
\begin{center}
\includegraphics[width=\textwidth]{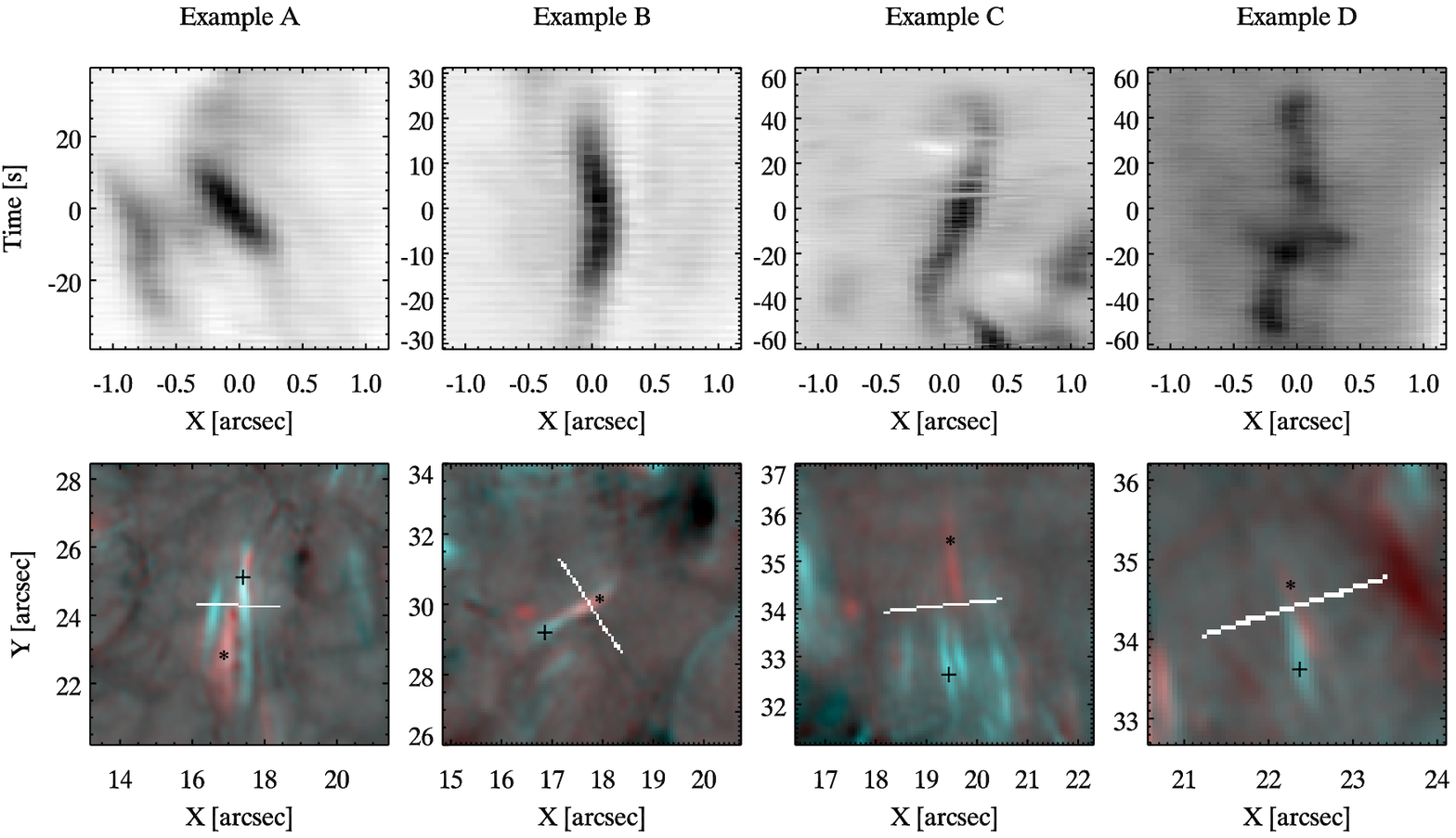}
\caption{Different transverse behaviours seen in RBEs: Example A has a motion in a single transverse direction, example B shows a single turn in the direction of the transverse motion, example C shows several turns in the direction of the transverse motion, example D is a number of RBEs repeating so close in space and time as to give the impression of being connected as a single longer lived RBE.  The shorter lived RBEs that make up an example such as example D can come from each of the other three groups.  Top row shows a spacetime diagram taken across the examples.  Bottom row shows a composite colour image of the examples where the first timeframe of the RBE is plotted in blue/green while the last timestep is plotted in red.  The line in the bottom row images signifies the slit from which the spacetime diagrams were made and the crosses and stars marks the approximate position of the RBEs in their first and last frame, respectively, to guide the eye.  Animations of the time evolution of these four examples are available as online material.}
\label{fig:examples}
\end{center}
\end{figure*}
 
Based on the number of transverse directional changes, RBEs can be grouped into different classes.  A natural first class (A) are all the RBEs that only display a transverse motion in one direction and hence, move diagonally in a spacetime diagram as seen in the first column of Fig.~\ref{fig:examples}.  For the second class (B) we took all the RBEs that turn once in their transverse motion, meaning all those that have a single change of direction in Fig.~\ref{fig:transverse}d.  An example of this behaviour is shown in the second column of Fig.~\ref{fig:examples}.
The remaining RBEs, the ones with several directional changes to their transverse motion in panel d of Fig.~\ref{fig:transverse}, are placed in the third class (C) and display the back and forth swaying motion of the example in the third column of Fig.~\ref{fig:examples}.  We see an increase of the average lifetime with an increase in the number of directional changes, i.e.\ class C has longer lifetimes than class B which has longer lifetimes than class A.

In addition to these three classifications, we found a fourth class (D) of RBEs.  These transcend the groups that are easily identified by Fig.~\ref{fig:transverse}d and are identified from the transverse spacetime diagrams as back and forth swaying regions which are very long lived and faint in contrast with very high contrast patches happening periodically along the faint region.  If these high contrast patches are taken as being the real RBEs, they can have a transverse motion which fits into any of the first three classifications.  In the right column of Fig.~\ref{fig:examples}, an event which is supposed to fit into class A is seen around the 0~s mark in the top panel.  This event seem to be a part of a longer lived event which is visible as a faint region which extends backwards and forwards in time compared to the group 1 event at 0~s and includes other small high contrast patches similar to the 0~s event.  

We provide accompanying movies as online material to vividly illustrate the different kinds of dynamical evolution patterns.  The movies are sequences of Dopplergram images centred on the RBE of interest at $(x,y)=(0,0)$.  The slit for the spacetime diagrams is positioned perpendicular to the main axis of the RBE.  We note that the high frequency jitter of background pattern is caused by remaining local misalignment, caused by seeing, between the sequentially recorded blue and red wings.  This slight misalignment is enhanced by the subtraction process when making the Dopplergrams.  Example A appears next to an already existing RBE on the left and extends quickly to its full length going towards the left hand side of the panel.  Looking more closely at this example reveals that the full length of the RBE is visible, but faint, before the apparent motion along the RBE coming from the footpoint, at the top of the FOV, has extended all the way to the maximum length.  The full extent of the RBE is progressively gaining higher contrast while the high contrast region propagates away from the footpoint until $t\approx15$~s after which the RBE rapidly fades.  For example B the movie shows an RBE which extends towards the bottom left corner.  In the middle of the lifetime of the RBE, the expansion of the RBE structure stops for a short while before continuing to expand.  Towards the end, the RBE stops again, and after a short while it starts to fade from the top down and vanish within a few timesteps (also see Fig.~\ref{fig:app_vel}b).  During its lifetime the RBE displays a small back and forth motion.  Example C appears at the bottom in the middle of the panel after a previous RBE has almost finished fading away.  It then continues to move upwards away from its footpoint at a nearly constant pace until, towards the end of of the lifetime, the top third vanishes while the remaining RBE structure continues its upward propagation.  During its lifetime the RBE also displays a sideways motion that moves mainly towards the right hand side of the panel.  This motion is, however, not constant and seems to start early in the RBE's lifetime and also come to a halt, or turning towards the left hand side near the end of the lifetime.  In the movie for example D, it becomes clear that there are four short-lived RBEs recurring in roughly the same location within a timespan of two minutes and that the reason for the rather wide second RBE, at t$\approx-15$~s, is due to a fifth short-lived RBE appearing right next to the second RBE.  Each of the four RBEs extend to their full lengths within a couple of seconds moving towards the top of the panel before fading from the FOV almost as rapidly.

\subsubsection{Wave Motion in RBEs}
\label{sect:wp_res}

\begin{figure}[htbp]
\begin{center}
\includegraphics[width=\columnwidth]{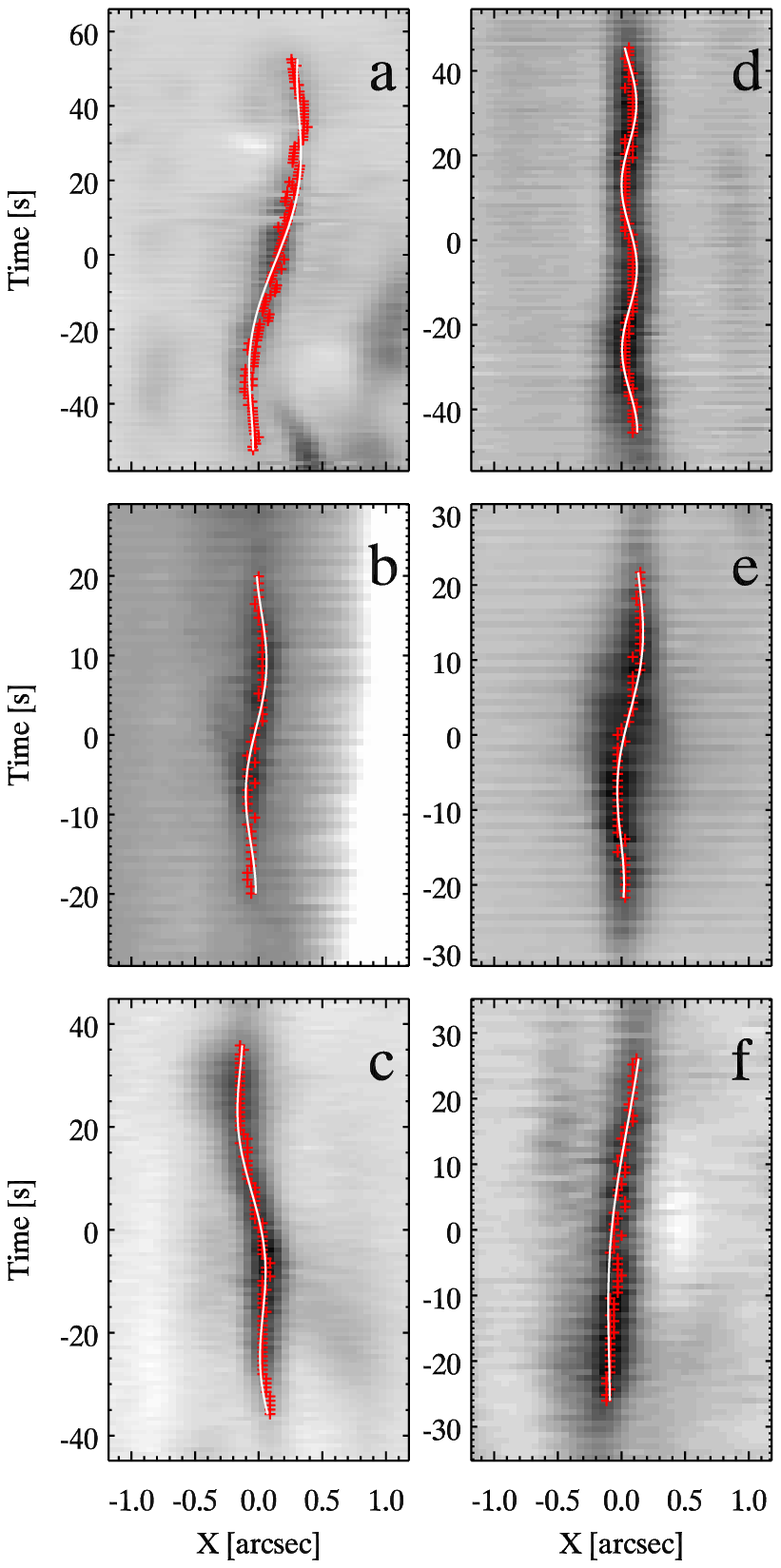}
\caption{Six examples showing a sinusoidal fit to the transverse motion of an RBE.  The red crosses mark the minimum intensity of the RBE at each timestep of the lifetime of the RBEs while the white line is the sinusoidal fit to the minimum intensity points.  The underlying images are spacetime diagrams from a slit which is across the RBEs and averaged over 20 pixels in the longitudinal direction of the RBEs.  Animations of the time evolution of these examples are available as online material.}
\label{fig:sinfit_ex}
\end{center}
\end{figure}

\begin{figure}[htbp]
\begin{center}
\includegraphics[width=\columnwidth]{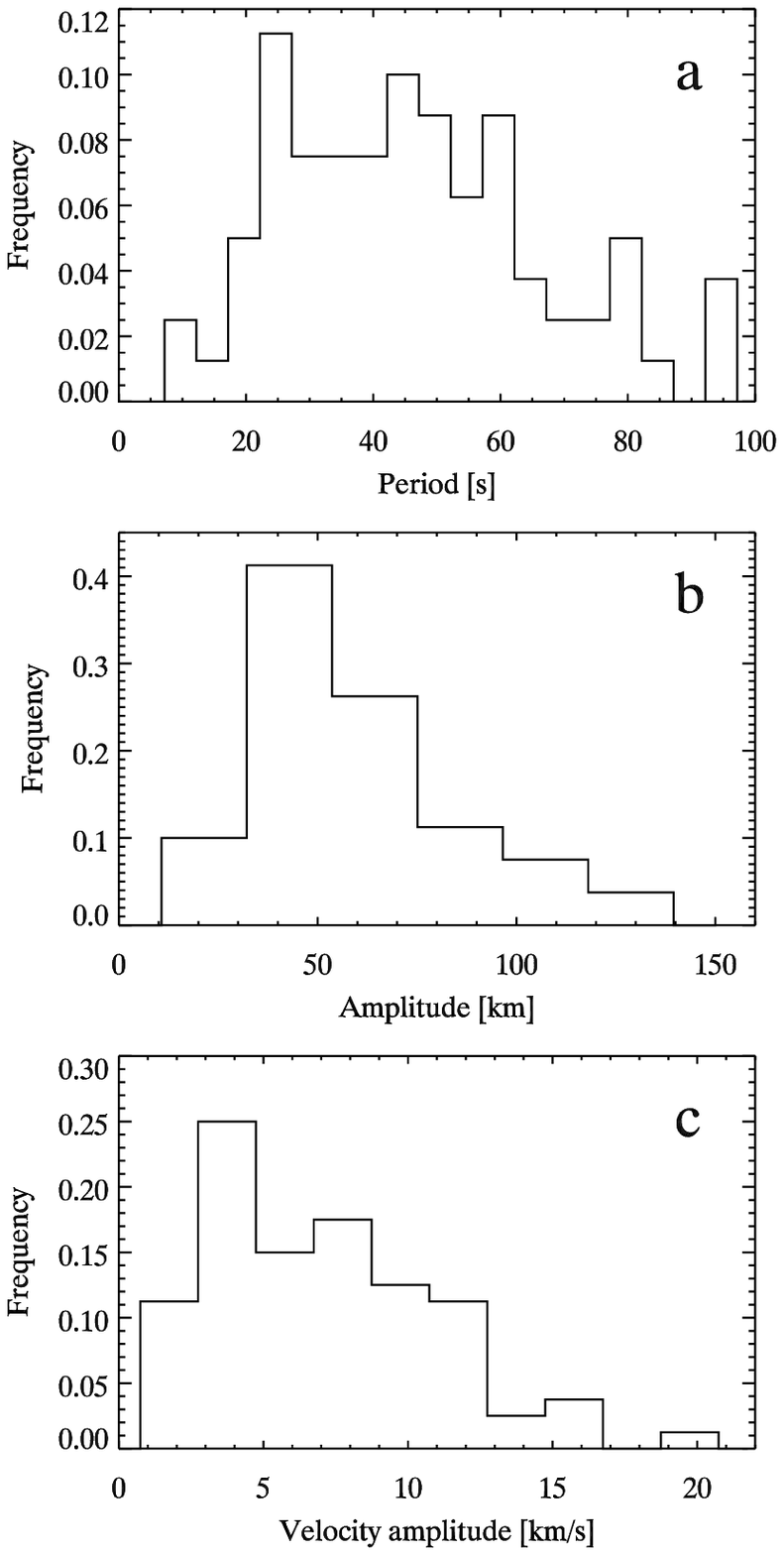}
\caption{Histograms of the properties of the sinusoidal fit to RBE transverse motions: (a) Periods, (b) amplitudes, (c) velocity amplitudes.}
\label{fig:sinfit_hist}
\end{center}
\end{figure}

From the grouping done in section \ref{sect:spatial_res}, all the RBEs in class C were investigated for sinusoidal waves in their transverse motion.  This was done by locating the lowest intensity pixel in the slit transverse to the RBE axis at each timestep and fitting a sinusoidal curve to these positions.  Out of 154 RBEs in class C, as many as 80 could be fitted to a sinusoidal function.  The other fits were not acceptable either because of contaminations in the transverse slit from nearby low intensity regions or irregular variations in the intensity of the RBE which ruined the sinusoidal fit. 

In Fig.~\ref{fig:sinfit_ex}, a selection of six such sinusoidal fits to the transverse motion are displayed, where the top left example is the same as example C in Fig.~\ref{fig:examples} and the other five are selected from the 80 accepted fits.  These fits allowed for a measure of the amplitudes and periods of the transverse motions, which are plotted in the histograms in Fig.~\ref{fig:sinfit_hist} along with a measure of the velocity amplitude.

The periods vary from 9.7~s up to 168~s with an average of 54~s, while the distribution peaks around 40~s.  For the amplitudes, the minimum lies at 21.5~km, which is exactly half the pixel scale of the dataset, while the maximum amplitude measured is 129~km with a clear peak of the distribution at 43~km.  The velocity amplitudes are calculated in the same way as was done for 
\citet{2011ApJ...736L..24O}, 
$v={2\pi\delta{x}}/P$ where $v$ is the velocity amplitude, $\delta{x}$ is the amplitude, and P is the period, and were found to lie between 1.5 and 21~\kms\ with a peak around 4~\kms\ and a median value of 7.5~\kms.  

For the accompanying movies to Fig.~\ref{fig:sinfit_ex}, the FOV has been rotated so that the RBE is oriented upwards.  From the accompanying movies the example in panel a, which is a zoomed-in version of the movie accompanying example C in Fig.~\ref{fig:examples}, displays an almost constant rise phase and a clear transverse motion during its lifetime.  Also, the shortening of the RBE structure when the top third vanishes can be seen clearly from this movie.  The example in panel b appears suddenly around t$\approx-20$~s over a length of approximately 2\arcsec.  The fading phase during the end of its lifetime appears to be from the top down and the sideways motion during the middle of its life is clearly discernible.  The example in panel c on the other hand extends rapidly from its footpoint and has several high contrast blobs propagating along the fully extended RBE.  This example also shows a clear back and forth sideways motion which from the spacetime diagram turns out to be sinusoidal.  In the movie for the panel d example we see a long lived RBE.  this RBE first extends to its full length before the entire RBE structure moves upwards.  The transverse back and forth motion is small but can be picked out from the movie.  The example in panel e grows very fast and reaches its full length within a few timesteps.  When examined more closely it appears as if sections of the RBE form simultaneously and progressively along the RBE axis creating the illusion of an extreme outwards propagation.  At the moment the full extent of the RBE has formed, t$\approx-20$~s, it displays a clear sinusoidal shape.  As it evolves in time we can see it display the wave motion that is inferred from the spacetime diagram.  The example in panel f displays a long period wave motion.  This RBE appears in place without any apparent propagation along the axis.  There is a small sideways motion seen in this RBE and it fades along its full length simultaneously.

\subsubsection{Longitudinal Velocities}
\label{sect:long_vel_res}

\begin{figure*}[t]
\begin{center}
\includegraphics[width=\textwidth]{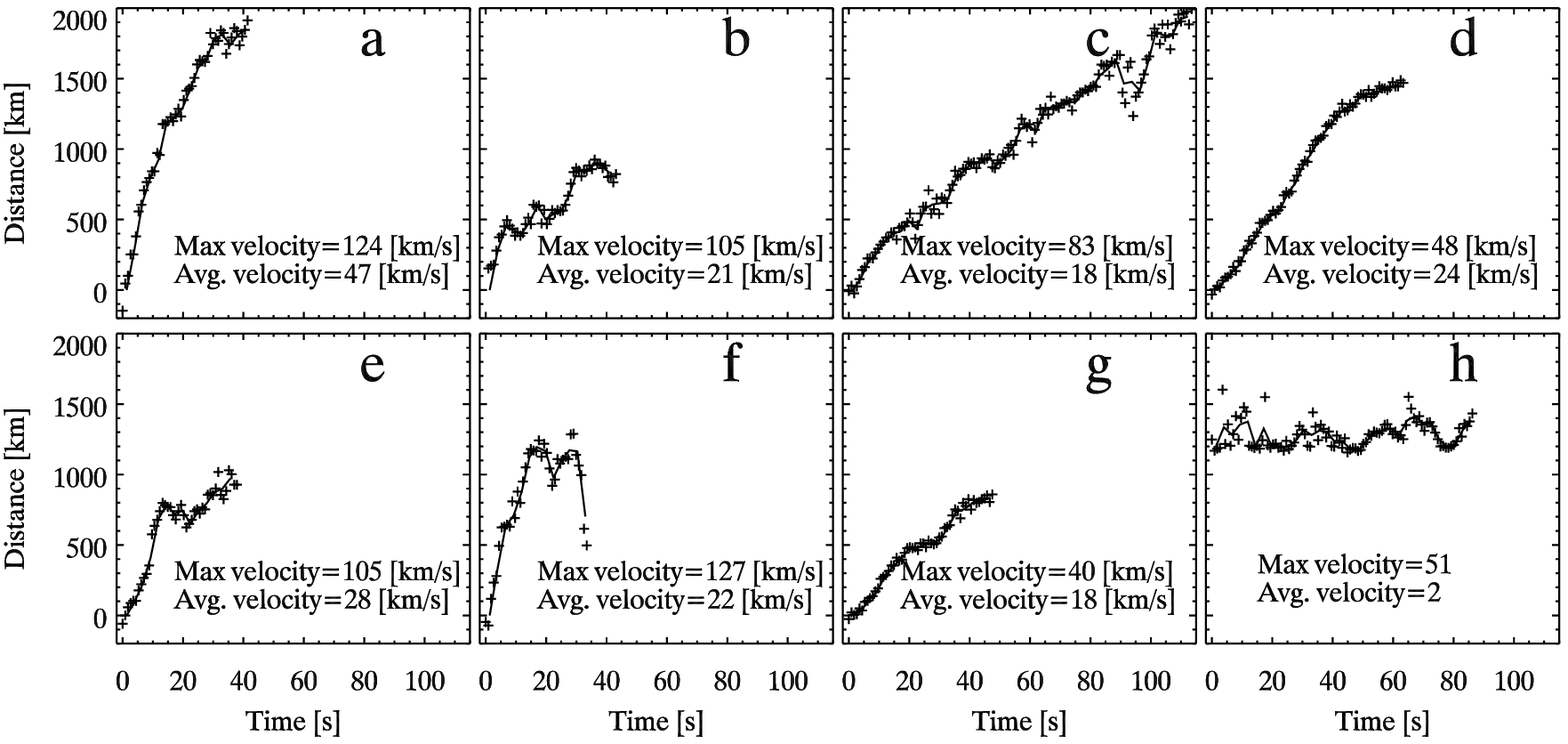}
\caption{The movement of the top of eight RBEs with respect to a fixed position close to the footpoint of the RBEs.  Animations of the time evolution of these examples are available as online material, examples a,b, and c accompany Fig.~\ref{fig:examples}.}
\label{fig:app_vel}
\end{center}
\end{figure*}

In Fig.~\ref{fig:app_vel} we plot the movement of the top of the RBE with respect to its footpoint for 8 different RBEs.  The first three examples on the top row are the same RBEs as in Fig.~\ref{fig:examples} A, B, C.  The next four examples were picked from the sample for their stable seeing which made them easier to measure, while the bottom right example is an RBE that appears to become visible along the full length of the RBE within a few seconds.  In the few examples studied more closely here, the largest movement of the top of an RBE with respect to its footpoint is about 2 Mm over a lifetime of about 120~s, which gives an average velocity of 18~\kms\ of the top point away from the footpoint.  The maximum and average velocities printed in the panels of Fig.~\ref{fig:app_vel} are calculated from the trajectory followed by the plotted line which is smoothed over 3 positions.

Accompanying movies to Fig.~\ref{fig:app_vel} are in the same format as for Fig.~\ref{fig:examples} and examples a, b, and c are discussed in Sect. \ref{sect:grouping_res}.  The panel d example moves towards the bottom of the panel in the movie while it is expanding in length in both the downward and upward direction.  Towards the end of its lifetime the RBE seems to split into two halves where the top half continues to move downwards and fades from the FOV.  The bottom half remains stationary for some time before continuing the motion towards the bottom of the panel and at the same time fading from the FOV\@.  In the movie for panel e we see an RBE extending towards the bottom from a footpoint close to the centre of the panel.  At a point towards the end of the lifetime of the RBE the top part vanishes and the rest of the RBE continues its movement towards the bottom of the panel.  It displays clear sideways motion resembling a sinusoidal wave pattern and fades from the FOV from the footpoint towards the top.  The example in panel f quickly reaches its full length extending from the footpoint upwards in the movie panel.  After reaching its maximum length the RBE lingers in this position for a moment before rapidly fading from the FOV giving the impression of an extreme downwards motion.  Panel g shows an example which goes in the direction from left to right.  When it appears it extends in both directions while keeping an overall motion to the right, away from the footpoint which lies to the left in the panel.  The fading of the RBE happens in the same way, with the top and bottom of the RBE fading simultaneously leaving the middle of the RBE as the last visible part.  This RBE appears to consist of several small concentrations.  The example displayed in panel h of Fig.~\ref{fig:app_vel} has its footpoint located towards the right edge of the panel in the accompanying movie.  In this RBE, the top half appears first and is fully formed within seconds.  The half closer to the footpoint then forms in the next few seconds, giving the impression of an apparent movement towards the footpoint.  When it fades away, this happens from the bottom up and the top part of the RBE is the last to vanish from the FOV.

\subsubsection{RBE Evolution in Simultaneous \CaIIIR-\Halpha\ Data}
\label{sect:wcf_res}

\begin{figure}[htbp]
\begin{center}
\includegraphics[width=\columnwidth]{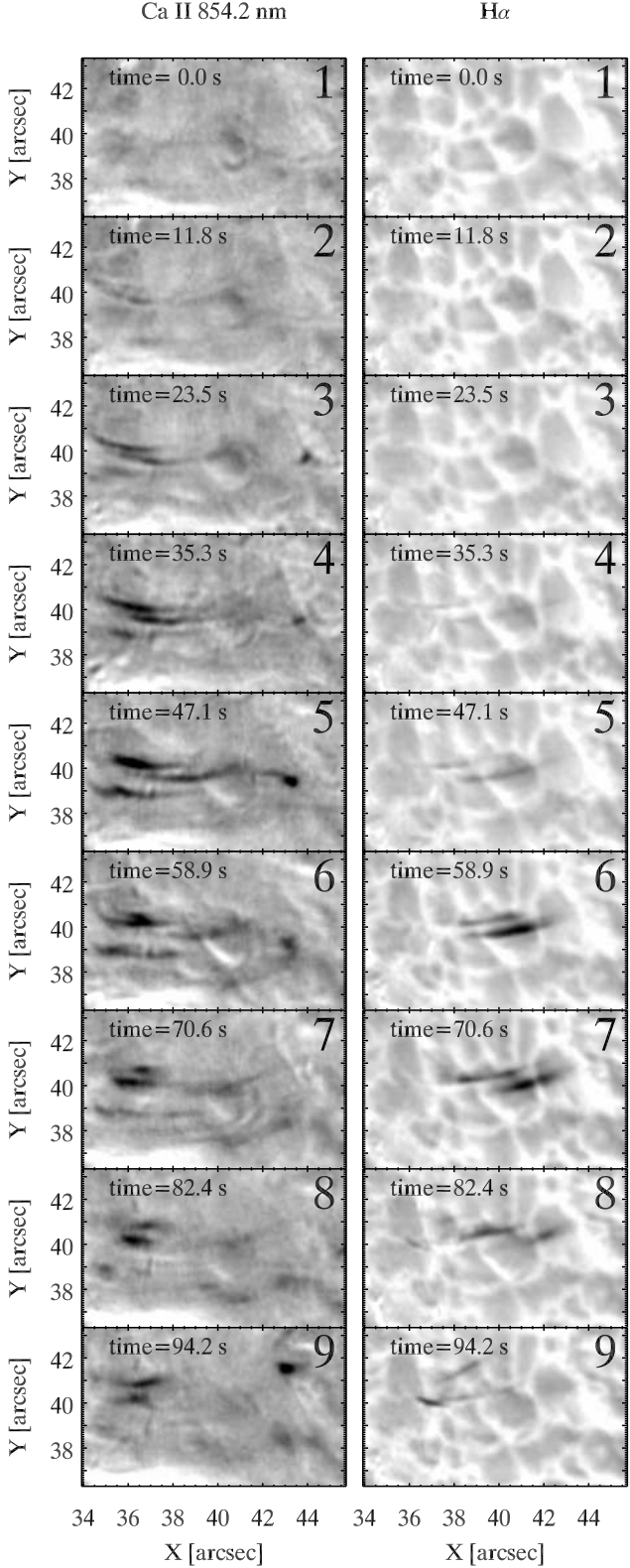}
\caption{An RBE found to appear in both \CaIIIR\ and \Halpha\ from the 27 June 2008 observations.  \CaIIIR\ images taken at the blue wing position of 20~\kms\ in the left column with their simultaneous counterpart \Halpha\ images from the blue wing position of 45~\kms\ in the right column.  An animation of this example is available as online material.}
\label{fig:evolution}
\end{center}
\end{figure}

In Paper~II, RBEs were detected in co-temporal \CaIIIR\ and \Halpha\ observations.  The RBEs were detected in single timesteps separately for the two spectral lines. Subsequently, the detected RBEs were automatically connected into multi-frame \CaIIIR\ and \Halpha\ RBEs.  Finally, for a number of these multi-frame RBEs, it was possible to connect the \CaIIIR\ and \Halpha\ chains.  For these linked multi-frame RBEs, it is possible to study the temporal evolution in two different spectral lines.

We find that for most of these automatically linked multi-frame RBEs, the RBE appears first in \CaIIIR\ or simultaneously in both lines.  Manual inspection of the few cases where the automated detection found the RBE in \Halpha\ first, revealed that there always is associated \CaIIIR\ signal, but too faint or small-scale to be identified by the automated method.

Figure \ref{fig:evolution} shows an example of the evolution of a multi-frame RBE from its initial detection in \CaIIIR\ in panel 3, left column, to when it fades from the FOV in \Halpha\ in panel 8, right column.  We focus on the RBE furthest to the right, centred on $(x,y) \approx (40,40)$ at $t=47$~s.  
The RBE in this example was not detected in \Halpha\ until panel 5, which is at least 20~s after the initial detection in \CaIIIR.  Also, the RBE is fading away from the FOV in \CaIIIR\ already in panel 6, and only barely visible in 7 and 8, whereas it is still strong in \Halpha\ for another 2 timesteps, $\sim$20~s. 

Figure \ref{fig:evolution} also shows very clearly that the \Halpha\ part of the RBE is located further away from the origin of the RBE when compared to the \CaIIIR\ part, with the origin being towards the left hand side of the images where the dataset contains a region of concentrated magnetic fields (see
Paper~II). 

\section{Discussion}

\subsection{Quiet Sun RBEs}

The studies of Papers~I~and~II have so far been conducted in coronal hole regions where the magnetic field is more unipolar.  In this study we have focused on a region with quiet Sun conditions, which are the conditions that on average cover 70$-$80\% of the solar surface 
\citep{1982QB539.M23P74...}. 

In
Paper~II, 
it was found that \Halpha\ RBEs in a coronal hole had Doppler velocities between 30 and 50~\kms.  We see from our quiet Sun observations that the range of Doppler velocities in our data match very well with what is seen in coronal holes, only shifted to lower values.

The Doppler widths seem to undergo the same effect as the Doppler velocities when moving from a coronal hole to quiet Sun.  While the range of Doppler widths are more or less the same as for the coronal hole in
Paper~II, 
it is shifted towards lower values.  These measurements of Doppler velocity and Doppler widths serve to characterise the spectral profiles of RBEs.  The Doppler velocity is related to the physical motion of plasma in the RBE and increased Doppler width could be related to enhanced temperatures or, alternatively, signs of unresolved substructures with varying velocities.  However, we caution against a direct physical interpretation of these measurements: only through detailed modelling of the radiation transport in the atmospheres of these structures knowledge on the exact physical conditions can be obtained.

While the Doppler velocities and widths are lower in quiet Sun than in a coronal hole, the physical lengths of RBEs do not seem to be very different for the two target area types.  The only differences are that we find slightly shorter RBEs in our quiet Sun data and a shift in the peak of the distribution to around 2.3~Mm instead of at 2.6~Mm, which can be attributed to the automated detection routine being allowed to search for these shorter events. 

It is possible that all these differences are naturally occurring as a result of the intrinsic differences in coronal hole and quiet Sun magnetic field configurations.  While coronal holes have predominantly unipolar fields and vertical field lines, quiet Sun harbours mixed polarity fields with more horizontal field lines and closed loops.  The difference in the magnetic field orientation might result in a different combined effect of the three types of spicule motion, field-aligned flow, transverse swaying, and torsional motion 
\citep{2012ApJ...752L..12D}.  
For example, for more inclined RBEs observed at disc centre, the field-aligned flow would only have a small component along the line-of-sight, so that the more varying effects of swaying and torsional motions could become more dominant in the observed Doppler properties.  In addition, it is possible that the different viewing angle, away from disc centre (viewing angle $\mu=0.84$) for the coronal hole of 
Paper~II 
and disc centre for the quiet Sun here, is contributing to the different values for the RBE properties in the two datasets.

Using the recurrence rate of one RBE per 500~seconds from 
\citet{2011Sci...331...55D}, 
\citet{2012arXiv1207.7048K} 
dismissed type II spicules as a sufficient source of hot plasma and energy to heat coronal loops and their importance for the temperature structure in the corona.  We now have datasets with a much higher cadence than before and we find average recurrence times around magnetic network where RBEs appear of 84~s and a minimum recurrence time of 4.3~seconds in the densest regions.  
Our results indicate that the conclusions of 
\citet{2012arXiv1207.7048K} 
are unwarranted, and also highlight the dangers of using simplified models and poorly determined observational characteristics. 

\vspace{1.5cm}
\subsection{Spatial Dependence of RBE Properties}
Figure \ref{fig:good_bad_ugly} shows three very different RBE examples of how the Doppler velocity and Doppler width changes with the position along the RBE.  In 
Paper~I, 
the large scale FOV picture was that RBEs in general would accelerate outwards from its footpoint and this view was confirmed by examining single RBEs which behaved similar to the first example in Fig.~\ref{fig:good_bad_ugly}.  In 
Paper~II, 
this general large scale FOV picture of having RBE Doppler velocities increasing outwards from their footpoints was confirmed, but it was noted that there exist a significant number of RBEs displaying a more erratic variation of their Doppler properties.  When making similar maps of the variation of Doppler properties along RBEs for our quiet Sun observations, we see a more complicated mixture of longitudinal variations (see Fig.~\ref{fig:rbe_fov}).  There are clear regions where the top of the RBEs are at a higher Doppler velocity and width than at the footpoint, but at the same time the exact opposite behaviour can be seen in other regions of the FOV, where the Doppler velocity and width are clearly lowest furthest away from the footpoints.  When examining the RBEs one by one we see that both of these behaviours are quite normal, and there are almost equal amounts of RBEs with Doppler velocities and widths increasing and decreasing away from the footpoints.  In addition to RBEs that show monotonic increasing and decreasing Doppler properties along their axis, there are others which display a much more complex configuration of Doppler velocities and widths.  These RBEs seem to have a change in their velocity in different locations along their lengths and may even display several changes both up and down in velocity in one single timeframe, as seen in the third example of Fig.~\ref{fig:good_bad_ugly}.  Although the Doppler velocity and width in general seem to be changing simultaneously and with the same trend, there are cases where the Doppler width changes more or less independently or oppositely of the Doppler velocity.  This is illustrated in Fig.~\ref{fig:qs_stats}d where the Doppler velocity vs. Doppler width has a looser correlation than for the coronal hole of 
Paper~II 
where there is a trend of larger Doppler width for increased Doppler velocity.

When examining the maps of Fig.~\ref{fig:rbe_fov}c and d, we see no apparent large scale trend in the spatial variations of the Doppler velocities and Doppler widths of RBEs.  At the same time, we note the absence of noise in the large scale pattern of Doppler variations.  If the Doppler velocity and width were completely independent of the position along the RBE, these maps would have a different appearance.  There is clearly more scatter in the large scale pattern than in the corresponding maps for the coronal hole targets of 
Paper~I and II 
(cf. Fig.~12 in Paper~I and Fig.~6 in Paper II), but the pattern in the colour distribution is very unlike it would be for a random distribution.    It is well possible that systematic trends in the variation of the Doppler properties along the RBE axis are hidden by the more complicated magnetic field geometry of the quiet Sun with the presence of more mixed polarity fields and associated more inclined and curved field lines.  In this less orderly configuration, large scale trends would be harder to discern in maps like Fig.~\ref{fig:rbe_fov}.  Moreover, we should realise that the combined effect of the three types of motions in spicules, field-aligned flow, swaying, and torsional motion, might be very different for varying angles between the line-of sight (LOS) and the magnetic field configuration.  Horizontal fields observed at disc centre could give rise to more of the torsional motions showing up in the RBE signal, whereas the more vertical field lines at upper chromospheric heights of a coronal hole may show more of the field-aligned motion or outflow of material.  We realise that the major unknown parameter in our analysis is the knowledge of the exact orientation of each RBE. Measurements of the magnetic field vector by means of various Stokes diagnostics would be a possible approach to become conclusive on the longitudinal variation of the Doppler properties of RBEs.

\subsection{RBE Dynamics from Temporally Resolved Data}
\subsubsection{Improved Lifetimes}

The first measurements of the lifetimes of type II spicules yielded values between 10$-$110~s
%
\citep{2007PASJ...59S.655D}, 
where the 10 second threshold was given by the cadence of the dataset in which this new type of spicule was discovered.  The extensive and systematic study of 
\citet{2012ApJ...759...18P} 
employed automatic detection on a larger sample of type II spicules and established a typical lifetime between 50$-$150~s with a peak in the distribution around 80~s.  For their disc counterparts, the lifetimes were estimated from a manual selection of 35 RBEs and measured to range between 10$-$70~s, with a few examples displaying lifetimes of up to 140 seconds 
(Paper~I).  
An automated method on a much larger sample (453), established a range 20$-$150~s 
(Paper~II) 
which was in agreement with the 
\citet{2012ApJ...759...18P} 
results.

For all these studies, the determination of the minimum lifetime of type II spicules was limited by the cadence of the observations.  Here, we resolve this limitation by analysing a much faster cadence dataset (0.88~s) and we make further improvements by employing a secondary manual measurement based on spacetime diagrams across the RBEs.  From Fig.~\ref{fig:lifetime} we see that when measuring the lifetimes in the transverse spacetime diagrams, we find that RBEs seem to have a lower limit to their lifetimes at approximately 5~s, which is much higher than the cadence of the dataset.  Lifetimes measured from the automated method alone, put the threshold down to the cadence of the dataset and closer evaluation concluded that this method was not reliable for this kind of measurements.  The lower threshold of lifetimes of RBEs at 5~s that is found here is likely the real minimum lifetime of RBEs in our dataset.  We find a range of lifetimes between 5 and 60~s with an average of 30~s, which indicated that datasets with a slow cadence will easily connect recurring RBEs in high density regions into longer lived RBEs.  We remark that these numbers should not be interpreted as "the" lifetime of RBEs/Type II spicules as we only have observations from one wing position.  The RBEs can be shifted out of the observed passband by acquiring higher or lower Doppler velocities due to field-aligned flows, swaying, and torsional motions, and continue their evolution out of range of the diagnostics.  This could be an explanation for the lower average lifetime we find for RBEs as compared to type II spicules at the limb.  The latter were measured in Hinode Ca~H intensity which is insensitive to these kind of changes in the Doppler shift.

\vspace{1.5cm}
\subsubsection{Transverse Motions}

In
Paper~I, 
manual measurements of 35 RBEs resulted in transverse velocities ranging from 0 to 20~\kms\ with an average of 8~\kms, whereas 
Paper~II, 
employing an automated method on a much larger sample for measuring multi frame RBEs, found a range 0-15~\kms\ with an average of 5~\kms.  By checking the 
Paper~I 
dataset with the automated detection method, 
Paper~II 
put the discrepancy of the transverse velocities down to the difference in cadence allowing for the detection of higher transverse velocities in the 
Paper~I 
dataset which had a cadence which was twice as fast as the 
Paper~II 
dataset.  It became clear that the temporal resolutions of these studies were not sufficient to be conclusive in these measurements which motivated us to pursue the acquisition of the high cadence dataset analysed here.

The average transverse velocity of RBEs in the high cadence dataset is comparable to the 
Paper~I 
result with a peak in the distribution which is comparable to 
Paper~II.  
What is different, however, is the increase in the distribution towards higher velocities, see Fig.~\ref{fig:transverse}c.  The distribution of the maximum transverse velocity in the lifetime of RBEs is also found to be shifted to higher velocities and is now in close agreement with the corresponding measurements for type II spicules at the limb by 
\citet{2012ApJ...759...18P} 
and the original measurements of 
\citet{2007Sci...318.1574D}.  

For the maximum displacement of the two extreme transverse positions we see a smaller difference between the high cadence data and 
Paper~II 
(see Fig.~\ref{fig:transverse}a).  There is an increase in the number of RBEs displaying maximum transverse displacements of 200~km and more, which is a direct result of having a higher cadence dataset where it is much easier to follow the RBEs with a very rapid transverse motion covering a larger distance.

We actually see a slight lowering of the integrated transverse displacement (Fig.~\ref{fig:transverse}b) in comparison with the lower cadence data of 
Paper~II.  
This can be attributed to the employment of manual measurements in our high cadence datasets compared to relying on the detected trajectories of the multi-frame RBEs in the low cadence dataset.  Only relying on the detected trajectories resulted in a significantly higher integrated displacement due to noise in the measurements.  The measurements of transverse displacement of RBEs is in close agreement with the corresponding measurements for type II spicules at the limb 
\citep{2012ApJ...759...18P, 
2007Sci...318.1574D}.  

 \subsubsection{Grouping of RBEs Based on Transverse Spacetime Diagrams}
 
With the extremely fast cadence of the 2011 May 05 dataset, it is possible to follow the transverse motion of RBEs in time.  From a manual inspection of all spacetime diagrams taken across the RBEs we saw a pattern of 3 different transverse behaviours, where the most common transverse motion is seen going in a single direction from one extreme position to the other within the lifetime of the RBE.  As the second most common transverse motion we found a single back and fort swaying motion which is characterised by one change in direction.  This behaviour is not necessarily going from one extreme through the other extreme and back again to its starting position, although that seems to be the case in example B in Fig.~\ref{fig:examples}.  The third behaviour of the transverse motion seen in our dataset can be classified as a proper back and forth swaying motion, where the direction of the transverse motion changes several times.  In Fig \ref{fig:examples} C, the RBE is seen to have a dominant transverse direction towards the right while exhibiting several turns in its transverse motion throughout its lifetime.  There are also examples that display a back and forth swaying motion while seemingly being centred around a middle point and hence, not displaying a dominant transverse direction.  Even though these groups appear to have different behaviour, these apparent differences could well be caused by different lifetimes and periods, and are thus not necessarily a sign that the different subclasses are caused by different mechanisms.

In addition to the three clearcut groups of RBEs, we also observed a fourth type in our transverse spacetime diagrams which consists of several repeating RBEs appearing so close in time and space as to give the impression of a longer lived RBE with concentrated blobs when examined in a spacetime diagram.  This fourth class, example D in Fig.~\ref{fig:examples}, supports the idea that the transverse swaying motions are in some sense independent of the formation of the RBEs.  That is, the waves are volume filling, with the RBEs occasionally tracing part the field lines that are affected.  This is further supported by the observation that the lifetimes of the different classes are on average increasing, i.e.\ we see longer lifetimes for class C than for class A which shows no change in direction.  Based on numerical simulations, 
\citep{2007Sci...318.1574D} 
proposed this scenario to interpret the swaying motions of type II spicules at the limb as tracers of Alfv\'{e}nic waves that permeate the chromosphere.  This scenario was later confirmed in AIA observations by
\citet{2011Natur.475..477M}.

\subsubsection{Wave Motion in RBEs}
Type II spicules were recently found to undergo three types of motion, a field aligned flow along the spicule axis, transverse swaying, and a torsional motion around its axis 
\citep{2012ApJ...752L..12D}.  
The transverse swaying in spicules was closely investigated by 
\citet{2011ApJ...736L..24O}, 
who concluded that they carry upward and downward propagating waves, as well as apparently standing waves with very high phase speeds.  We find that for part of the RBEs of group C, the sideways motion can be fitted by a sinusoidal function (Fig.~\ref{fig:sinfit_ex}).  The periods and amplitudes of the these sinusoidal waves (Fig.~\ref{fig:sinfit_hist}) correspond well to those found in type II spicules 
\citep{2011ApJ...736L..24O}.  

The study of transverse motions in RBEs done by 
\citet{2012arXiv1207.6417Y} 
also find a sinusoidal behaviour.  With periods averaging 90~s and amplitudes of 200~km, their results are not directly comparable to those found here, but with a cadence of 10~s it is likely that the RBEs studied were of the more extreme type when it comes to at least lifetimes and possibly also transverse movement.  Because of the lower cadence of their dataset it cannot be excluded that the long lived RBEs observed are several distinct RBEs identified as one due to the superposition in time.  However, it is clear that type II spicules and RBEs both display transverse sinusoidal wave motion.

\vspace{1cm}
\subsubsection{Longitudinal Velocities}
Type II spicules have high apparent rising velocities, 50--150~\kms 
\citep{2007PASJ...59S.655D}. 
\citet{2012ApJ...759...18P} 
examined type II spicules in both coronal holes and quiet Sun regions, and found the maximum velocities of these spicules to typically lie between 40 and 100~\kms, with a few cases going beyond 150~\kms.  They also found that the quiet Sun spicules had a lower maximum velocity on average and a sharper drop-off after 100~\kms\ than coronal hole spicules.

We tracked the movement of the top of  a number of RBEs and found maximum velocities from approximately 40 to 125~\kms.  We also see that the top of an RBE does not move away from the footpoint linearly nor exponentially, which would be expected if the RBEs would have a constant or increasing velocity as it propagates away from the footpoint.  In fact, the behaviour of the top part of the RBE is somewhat erratic and complex, with the RBE seemingly coming to a halt or even retreating for a short period of time before continuing its motion away from the footpoint.  This can not be explained by a relatively simple (in terms of spatio-temporal variation) mass motion along an outwardly expanding jet alone.  Full understanding of such complex behaviour will likely have to wait for detailed theoretical modelling of these jets.  However, it is clear that the presence of time-varying torsional disturbances that propagate at high speeds can play an important role in understanding complex apparent motions seen in observations at a fixed wavelength.  Given the uncertainty and variations across the FOV of angle between the magnetic field and LOS, our Doppler observations are likely sensitive to a varying mix of field-aligned motions and torsional motions.  The combination could well provide an explanation for the often erratic apparent motions observed by means of Doppler diagnostics in the wings of the \Halpha\ and \CaIIIR\ spectral lines.  The presence of very high phase speeds, as detected by 
\citet{2011ApJ...736L..24O} 
and caused by mostly standing waves, can then provide a natural explanation for the sometimes sudden appearance of an RBE over its whole length within a few timesteps.

In 1~s cadence red wing \Halpha\ IBIS observations of a region near the limb, 
\citet{2012ApJ...755L..11J} 
report chromospheric thread-like structures appearing and disappearing along their entire length within a few seconds and seemingly out of nowhere.  They interpret this observation as indirect evidence for the existence of plasma sheets in the solar chromosphere 
\citep[cf.][]{2011ApJ...730L...4J}.  
We confirm the observation of sudden appearance for some RBEs in our data, see Fig.~\ref{fig:app_vel}h.  In their complex spectrum of dynamical behaviour, we see some RBEs appear as thin threads without a clear rise phase in a few timesteps.  In contrast to 
\citet{2012ApJ...755L..11J}, 
we interpret this sudden appearance in the context of torsional modulation and high phase speeds of standing waves in spicules 
\citep[as conclusively shown by][]{2011ApJ...736L..24O}, 
rather than the superposition effects in two-dimensional sheet-like structures.

\subsubsection{RBE Evolution in Simultaneous \CaIIIR-\Halpha\ Data}

In 
Paper~II, 
it was found that \CaIIIR\ RBEs are located closer to the anticipated footpoints in the magnetic network regions than \Halpha\ RBEs and that \Halpha\ RBEs are a continuation of the \CaIIIR\ RBEs.  Here, we extended the analysis of the 2010 June 27 dataset to study the \CaIIIR-\Halpha\ connection in the temporal evolution of RBEs.  We found that for all the RBEs detected in both spectral lines, the RBE appeared first in \CaIIIR\ or simultaneously in both lines.  We could not identify cases of an RBE appearing first in \Halpha\ without associated RBE signal in \CaIIIR.  

The observation that some RBEs appear first in \CaIIIR\ before becoming visible in \Halpha, suggests that for these RBEs the origin of the jet is located in the lower chromosphere, below the sampling heights of the \Halpha\ line.  For the RBEs appearing in both lines simultaneously, the origin must be located relatively higher in the atmosphere.

In the example in  Fig.~\ref{fig:evolution} we see that the \CaIIIR\ part of an RBE appears at least 20~s before the RBE becomes visible in \Halpha.  It also vanishes from the FOV in \CaIIIR\ earlier than in \Halpha.  For this case, \CaIIIR\ shows the RBE closer to its origin and early in its life, while \Halpha\ shows us the later stages of the RBE's life, further away from its footpoint and apparently at greater height.  Progressively fading first in \CaIIIR\ before \Halpha\ could also be related to heating of the spicule.  This observation is interesting in connection with the work of 
\citet{2011Sci...331...55D} 
who find that the fading of Hinode Ca~H spicules is directly linked to features appearing at transition region temperatures in He~II~304~\AA\ and sometimes even appearing in coronal diagnostics.

\section{Conclusion}
We addressed various aspects of the temporal evolution of RBEs by analysing a number of high quality datasets: a dual line \Halpha+\CaIIIR\ timeseries, a spectrally resolved \Halpha\ series, and a coarsely sampled, extremely high-cadence series.  Our analysis adds more weight to the interpretation of RBEs being the disc counterpart of type II spicules, for example from the close agreement in transverse speed and displacement when compared to type II spicules at the limb 
\citep{2012ApJ...759...18P} 
 and from the agreement between sinusoidal wave parameters in RBEs we find here and those found in type II spicules at the limb 
\citep{2011ApJ...736L..24O}.  
At the same time, the picture of the dynamical behaviour of RBEs and spicules that arises from these kind of datasets is a very complex one.  Even in the extremely high-cadence timeseries, specially tailored to resolve the dynamics at the shortest time scales, the variation in temporal behaviour is bewildering.  We see RBEs move sideways with velocities up to 55~\kms, while others get hardly displaced at all.  Some RBEs live as short as 5~s, while others live for more than 2~minutes.  Some RBEs have a distinct rise phase with velocities up to 127~\kms, some rise with velocities below 40~\kms, and others appear suddenly along the full length within a few seconds without a distinct rise.  In dense regions, we see RBEs appear in rapid succession, sometimes as a short-lived apparent densification of a faint background structure.

Now that spectroscopic data of spicules at the limb have established that spicule dynamics is governed by three different types of motion, field-aligned flows, transverse swaying, and torsional motion 
\citep{2012ApJ...752L..12D}, 
we realise that the complex appearance of RBEs has to be interpreted within this threefold dynamical framework.  Taking into account these three different types of motion, and the observed presence of apparently standing waves in type II spicules, the sudden appearance of RBE features at extremely high phase speeds is no longer as difficult to explain as suggested by 
\citet{2012ApJ...755L..11J} 
(who have used this observational finding to propose a sheet-like nature of spicules).  While the extreme cadence of the fast dataset was useful to resolve RBEs temporally, it is clear that the sparse spectral sampling is insufficient to fully characterise the RBEs evolution.  Significant changes in Doppler shift due to swaying or torsional modulation require denser spectral coverage.  The measurement of a minimum lifetime of 5~s at one fixed \Halpha\ wavelength indicates that the observer has at least a few seconds to build up a more extended spectral sampling, so the characterisation of RBE Doppler modulation appears to be within reach of present-day instrumentation.

\acknowledgements
The Swedish 1-m Solar Telescope is operated by the Institute for Solar Physics of the Royal Swedish Academy of Sciences in the Spanish Observatorio del Roque de los Muchachos of the Instituto de Astrof\'{\i}sica de Canarias.  
B.D.P was supported through NASA contracts NNG09FA40K (IRIS), NNX11AN98G, and NNM12AH46G.  
We thank Jaime de la Cruz Rodr{\'i}guez and Eamon Scullion for their help during the observations.
This research has made use of NASA's Astrophysical Data Systems.

\end{document}